\documentclass[onecolumn,letter]{jpsj3}
\usepackage{txfonts}
\usepackage{mathtools}
\usepackage{braket}
\usepackage{color}

\title{Quantum Thermal Hall Effect in a Time-Reversal-Symmetry-Broken Topological Superconductor in Two Dimensions : Approach From Bulk Calculations
}

\author{Hiroaki Sumiyoshi \thanks{E-mail:h.sumiyoshi@scphys.kyoto-u.ac.jp } and Satoshi Fujimoto}
\inst{Department of Physics, Kyoto University, Kyoto 606-8502, Japan} 

\abst{We discuss thermal transport of two-dimensional topological superconductors (TSCs) with broken time reversal symmetry,
which are described by Bogoliubov-de Gennes (BdG) Hamiltonians. 
From the calculations of bulk quantities only, without refereeing to Majorana edge states,
we show that the thermal Hall conductivity of two-dimensional TSCs in the low-temperature limit
is quantized in multiples of $\frac{1}{2}\frac{\pi T}{6}$, which is exactly one half of the value of quantization in the case of the integer quantum Hall effect,
and that this exact half-quantization is caused by the structure of the Nambu spinor and the particle-hole symmetry,
which BdG Hamiltonians generally have.
In the case of spinless chiral $p$-wave superconductors, this result is in perfect agreement with the argument based on the Ising conformal field theory with the central charge $c=1/2$, which is an effective low-energy theory of the Majorana edge states. }

\kword{thermal Hall effect, topological superconductor}

\begin{document}
\maketitle
{\it Introduction} --- Topological insulators (TIs) and superconductors (TSCs) are new quantum phases
which are characterized by bulk topological invariants.
It has been found that some of these invariants directly characterize the transport properties of the system.
For example, the integer quantum Hall effect (IQHE) state is characterized by the topological number, the TKNN (or first Chern) number $C_1$,
and this invariant appears in the Hall conductivity $\sigma_{xy}$ in the low-temperature limit: $\sigma_{xy}=C_{1}e^2/h$,
which has been derived from the linear response theory.\cite{tknn,kh}
A similar idea is also applicable to topological superconductors without time reversal symmetry in two dimensions
(i.e. class C and D in the topological periodic table\cite{srfl}),
for which topological nontriviality is characterized by the nonzero TKNN number.
However, in this case, it is not charge transport but thermal transport that characterizes the topological feature, 
since charge is not conserved, while energy is still a conserved quantity in superconducting states. 
For instance, in the case of  spinless chiral $p$-wave superconductors,
which have the TKNN number $C_1=\pm 1$, \cite{vol}
Read and Green showed that by using the effective low-energy theory
for Majorana edge states,
the thermal Hall coefficient in the low-temperature limit is precisely given by $c\frac{\pi T}{6}$
 (in the unit of $k_{\rm B}=1$, $\hbar=1$),
where $c=1/2$ is the central charge of the Ising conformal field theory (CFT) which describes 
the Majorana edge state.\cite{rg,nomura}
Note that this value is one half of the value of IQHE states, which have edge modes
described by the chiral Luttinger liquid theory, i.e. CFT with the central charge $c=1$.\cite{rg}
The Read-Green's argument is based on the CFT for the edge states, and the transport coefficients
are not calculated from the bulk Hamiltonian.
This means that the bulk-edge correspondence is ensured only by the existence of the underlying effective low-energy topological field  theory, i.e. the gravitational Chern-Simons theory for thermal responses.\cite{rg}
However, the CFT for the edge state can not predict the sign of the Hall conductivity, though its magnitude up to the sign
is correctly determined. 
For instance, the above-mentioned formula $c\frac{\pi T}{6}$ is derived by assuming that its sign is positive.
Actually, the sign of the Hall conductivity depends on the bulk band structure.
Also, it is highly challenging to derive the gravitational Chern-Simons theory from microscopic Hamiltonians
of solid state systems.
Thus, it is desirable to obtain the quantum thermal Hall conductivity directly from the bulk microscopic Hamiltonian
without referring to the edge theory. 
In fact, the calculation of the quantized thermal Hall conductivity from the bulk Hamiltonian is highly nontrivial 
even for the case of the Chern insulator (or the IQHE state), as was noticed by Qin, Niu, and Shi recently.\cite{enemag}
They pointed out that to obtain quantized thermal Hall conductivity which is associated with the TKNN number,
one needs to extract the contributions of the energy magnetization current, since the magnetization current does not
participate into the transport current induced by applied temperature gradient.
Their procedure successfully gives the correct expression of the thermal Hall conductivity for the Chern insulator $\kappa_{xy}=\frac{\pi T}{6}C_1$ at sufficiently low temperatures.
Also this result is in agreement with the edge theory of the Chern insulator, i.e. CFT with the central charge $c=1$ :
if there are $n$-channel of the edge modes, the edge theory gives $\kappa_{xy}=c\frac{\pi T}{6}n$ which coincides
with the bulk result up to the sign, implying $|C_1|=cn=n$. 

Recently, Nomura et al. clarified the relation between the bulk TKNN number and the half-quantized thermal Hall conductivity in the case
of two-dimensional (2D) Majorana fermions on the surface of three-dimensional TSCs, which is related to 2D spinless $p+ip$ superconductors.\cite{nomura,comment2}
However, for more general cases of 2D TSCs such as chiral $d$-wave superconductors and
$s$-wave superconductors with the Rashba spin-orbit interaction (SOI), \cite{stf1,sf,stf2}
it has not been well understood how we can associate their TKNN number, which is an integer, 
with the CFT prediction mentioned above; i.e. the thermal Hall conductivity is not integer-multiple of
$\frac{\pi T}{6}$, but half-quantized.
To solve this issue, in this letter, we microscopically derive the thermal Hall conductivity from the bulk Hamiltonian, without using the edge theory.
There are two important key points for the microscopic calculation of the thermal Hall conductivity:
(i) we should take into account the energy conservation law properly, and (ii) the contributions from energy magnetization
must be extracted to obtain transport currents as pointed out by Qin, Niu, and Shi.\cite{enemag}
For this purpose, we, first, clarify the symmetric properties which Bogoliubov-de Gennes (BdG) Hamiltonians generally possess.
Next, we define the energy current operator which preserves the continuity equation.
Finally, we calculate the thermal Hall coefficients of TSCs by introducing gravitational fields which act as mechanical forces inducing heat currents, and by using the Kubo formula with corrections
from energy magnetization contributions.\cite{enemag} 
It is shown that the thermal Hall conductivity is quantized in the low-temperature limit
 exactly to one half of the value in the case of Chern insulators and IQHE states.

{\it Model, symmetry, and preliminaries} --- In this section, we introduce a model which describes 2D topological superconductors 
with broken time-reversal symmetry.
The argument developed in this letter is applicable to large classes of 2D topological superconductors such as
chiral $p$-wave superconductors,\cite{rg} chiral $d$-wave superconductors, and $s$-wave superconductors with the 
Rashba SOI under magnetic fields.\cite{stf1,sf,stf2}

The Hamiltonian we start with is given by
\begin{equation}
\hat{\mathcal{H}}\equiv \int d^d r \hat{\psi} ^ \dagger (r) \frac{1}{2}  H_{BdG} (r,\partial_r) \hat{\psi} (r),
\end{equation}
where each operator has indices $(a,s)$.
$a(=\pm1)$ is an index of the Nambu space and $s$ is a spin index.
$\hat{\psi}_{-1s}(r)$ and $\hat{\psi}_{1s}(r)$ are defined as $\hat{c}_s(r)$ and $\hat{c}^{\dagger}_s(r)$, respectively,
where $\hat{c}_s(r)$ ($\hat{c}_s^{\dagger}(r)$) is an annihilation (creation) operator of an electron,
and the BdG Hamiltionian $H_{BdG}$ generally takes the form:\cite{bdg,srfl}

\begin{eqnarray}
H_{BdG \,asa's'}(r,\partial_r) \equiv \left( 
\begin{array}{ccc}
&a'=-1&a'=+1\\
a=-1 &K_{ss'} (r,\partial_r)& -\Delta _{ss'}(r,\partial_r)  \\
a=+1 &\Delta _{ss'}^*(r,\partial_r)& -K_{ss'} ^*(r,\partial_r)
\end{array}
\right),
\end{eqnarray}
where $K_{ss'} (r,\partial_r)$ and $\Delta _{ss'}(r,\partial_r)$ are operators including differential operators and functions of $r$.
In the case where $K(r,\partial_r)=-\frac{1}{2m}\nabla^2+V(r)-\mu$ and $\Delta (r,\partial_r)=-\hat{\Delta}(-i\partial _x-\partial_y)$
[$V(r)$ is a potential energy and $\hat{\Delta}$ is a complex constant],
 the model describes spinless chiral $p$-wave superconductors \cite{rg}.
 On the other hand, in the case where $K(r,\partial_r)_{ss'}=\{ (-\frac{1}{2m}\nabla^2+V(r))\delta_{ss'}-h\sigma_z + \boldsymbol{\sigma} \cdot \mathbf{g}(\partial _r) \}_{ss'}$
and $\Delta_{ss'}(r,\partial_r)=(-i \Delta \sigma_y)_{ss'}$ [$h$ is a real constant, $\Delta$ is a complex constant, $\boldsymbol{\sigma}=(\sigma_x,\sigma_y,\sigma_z)$ are Pauli matrices, and $\mathbf{g}(\partial_r)=2\lambda(-i\partial_y,i\partial_x,0)$],
the model describes $s$-wave superconductor
with the Rashba spin-orbit interaction (SOI).\cite{stf1,sf,stf2}

For the derivation of the formula for the thermal Hall conductivity, it is useful to
clarify the symmetry of our system. 
The BdG Hamiltonian is Hermitian and preserves the particle-hole symmetry (PHS): $H_{BdG\,as\,a's'}=-H_{BdG\,-as\,-a's'}^{*}$.
Consequently, if $f_{as}(r)$ is the solution for the eigenvalue $E$, $f^*_{-as}(r)$ is the solution for the eigenvalue $-E$.
As a result, if the system is periodic, the eigenfunctions for $H_{BdG}$ can be written as the form $f_{kn\,as}(r)=e^{ikr}u_{kn\,as}(r)$,
where $u_{kn\,as}(r)$ has the same periodicity as the system,
and  the eigenenergys and eigenfunctions can preserve the symmetries
\begin{eqnarray}
E_{kn}=-E_{-k-n}, \quad f_{kn \, as}(r)=f^*_{-k-n\,-as}(r). \label{sym}
\end{eqnarray}
We take $E_{nk}\,(n=1,2,3\cdots)$ positive and $E_{nk}\,(n=-1,-2,-3\cdots)$ negative, and normalize the eigenfuctions so that $\sum_{as}\int d^d r f^*_{kn\,as}(r)f_{kn\,as}(r)=1$.

Now, we obtain the expressions for correlation functions of creation and annihilation operators of Bogoliubov quasiparticles, which are utilized in the following calculations.
We introduce annihilation (creation) operators of Bogoliubov quasiparticles:
$\hat{\psi}_{kn}\equiv \sum_{as}\int d^d rf^*_{kn\,as}(r) \hat{\psi}_{as}(r)$.
By using the symmetry eq. (\ref{sym}), we can derive the symmetry of these operators $\hat{\psi}_{kn}=\hat{\psi}^{\dagger}_{-k-n}$
and the anticommutation relations:
$\left\{\hat{ \psi}_{kn},\hat{\psi}^{\dagger}_{k'n'} \right\}=\delta_{kk'}\delta_{nn'}, \quad
\left\{ \hat{\psi}_{kn},\hat{\psi}_{k'n'} \right\}=\left\{ \hat{\psi}^{\dagger}_{kn}, \hat{\psi}^{\dagger}_{k'n'} \right\}=\delta_{k-k'}\delta_{n-n'}$.
 (Note that the latter is nonzero.)
Since the Hamiltonian can be transformed into $\hat{\mathcal{H}}=\frac{1}{2}\sum_{kn} E_{kn} \hat{\psi}^{\dagger}_{kn} \hat{\psi}_{kn}
=\sum_{kn,n>0}E_{kn}\hat{ \psi}^{\dagger}_{kn}\hat{\psi}_{kn}+(\mathrm{c-number})$,
we get the expressions for correlation functions:
\begin{align*}
&&\langle  \hat{\psi}_{kn}\hat{\psi}_{k'n'}\rangle&=f(E_{k'n'})\delta_{n-n'}\delta_{k-k'}= (1-f(E_{kn}))\delta_{n-n'}\delta_{k-k'},\\
&&\langle  \hat{\psi}_{kn}^{\dagger}\hat{\psi}_{k'n'}\rangle&=f(E_{k'n'})\delta_{nn'}\delta_{kk'}, \\
&&\langle  \hat{\psi}_{kn}\hat{\psi}^{\dagger}_{k'n'}\rangle&=(1-f(E_{k'n'}))\delta_{nn'}\delta_{kk'}, \\
&&\langle  \hat{\psi}_{kn}^{\dagger}\hat{\psi}_{k'n'}^{\dagger} \rangle&=(1-f(E_{k'n'}))\delta_{n-n'}\delta_{k-k'} =f(E_{kn})\delta_{n-n'}\delta_{k-k'},
\end{align*}
where the function $f(E)=1/(e^{\beta E}+1)$ is the Fermi distribution function at the temperature $T=1/\beta$ and
the chemical potential $\mu=0$,
and $\langle \hat{X} \rangle$ is the the statical average of the operator $\hat{X}$:
$\langle \hat{X} \rangle \equiv \mathrm{Tr}\left[ e^{-\beta \hat{\mathcal{H}}} \hat{X} \right]/\mathrm{Tr}\left[ e^{-\beta \hat{\mathcal{H}}}\right]$.
Applying Wick's theorem, we get the expression for the four-point correlation function:
\begin{align}
&\langle \hat{\psi}^{\dagger}_{k_1n_1}  \hat{\psi}_{k_2n_2} \hat{\psi}^{\dagger}_{k'_{1}n' _{1}}  \hat{\psi}_{k'_{2}n'_{2}} \rangle 
=f(E_2)f(E_{2'})\delta_{12}\delta_{1'2'}\nonumber\\
&+f(E_{1})(1-f(E_{2}))(\delta_{12'}\delta_{21'} -\delta_{1-1'}\delta_{2-2'} ),\label{4p}
\end{align}
where $E_{1}$, $\delta_{12}$ and $\delta_{1-2}$ are the abbreviations for 
$E_{k_{1}n_{1}}$, $\delta_{k_{1}k_{2}}\delta_{n_{1}n_{2}}$
and $\delta_{k_{1}-k_{2}}\delta_{n_{1}-n_{2}}$ respectively.
Note that the last term of (\ref{4p}), $-\delta_{1-1'}\delta_{2-2'}$, does not appear
in the four-point correlation function of  the systems of ordinary fermions,
and is inherent for the Nambu spinor.
This term plays an important role for the derivation of the half-quantized thermal Hall conductivity
as shown in the following.

{\it Energy current operator} ---
In general, for the microscopic argument on transport phenomena,
it is important to define correctly the "charge"  and "current" field operators which satisfy the continuity equation.
Therefore, in this section, to consider thermal transport,
we define the energy density operator and the energy current operator.
To apply the linear response theory, we introduce a gravitational field which gives rise to mechanical
forces inducing heat current flow.\cite{lu}

First, we, for simplicity, restrict the form of the BdG Hamiltonian to
\begin{align}
H_{BdG}(r,\partial_r)=\sum_{ij}A_{ij} \partial _i \partial_j+\sum_{i} \{ B_{i}(r)i\partial_i+i\partial_{i}B_{i}^{\dagger}(r) \}+C(r).
\label{Hbdg}
\end{align}
$A_{ij}$ is a constant Hermitian [i.e. $(A^{\dagger}_{ij})_{as\,a's'}\equiv (A_{ij})^{*}_{a's'\,as}=(A_{ij})_{as\,a's'}$] matrix which preserves $A_{ij}=A_{ji}$,
and the first term includes the kinetic energy and the superconducting gap which includes second-order differential operators.
$C(r)$ is a Hermitian-matrix-valued function of $r$,
and includes  the periodic potentials, the superconducting gap, the Zeeman energy, and so on.
$B_{i}(r)$ is a matrix-valued function of $r$ which is raised by gauge potentials, the spin-orbit interactions,
or superconducting gap which includes first-order differential operators.
Now we can get the equation:
$\sum_{i} \{ B_{i}(r)i\partial_i+i\partial_{i}B_{i}^{\dagger}(r) \}=
\sum_{i} \{ \tilde{B}_{i}(r)i\partial_i+i\partial_{i}\tilde{B_{i}}(r))+[\partial_i,\tilde{B'}_{i}(r)] \}$,
where $\tilde{B}_{i}(r)$ and $\tilde{B'}_{i}(r)$ are Hermitian matrices: $B_{i}(r)=\tilde{B}_{i}(r)+i\tilde{B'}_{i}(r)$,
and the last term of the RHS of the equation 
is noting but a Hermitian matrix, so
it can be absorbed into $C(r)$.
Therefore, we take $B_{i}(r)$ Hermitian.
Note that the Hamiltonians of the models mentioned above, the models of chiral $p$-wave (or $d$-wave) superconductors and $s$-wave superconductors with the Rashba SOI, are expressed in the form eq. (\ref{Hbdg}). From the the Hamiltonian eq. (\ref{Hbdg}),
we define the velocity operator :
\begin{align}
v_i\equiv i \left[ H_{BdG},r_i \right]. \label{vel2}
\end{align}
 $v_i$ is Hermitian and preserves PHS: $v_{i\, asa's'}=+v^{*}_{i\, -as-a's'}$.
(Note the sign "$+$" )

Next, we define the energy density operator and the energy current operator.
By using the method of integration by parts, the Hamiltonian $\hat{\mathcal{H}}$ can be rewritten as
\begin{align*}
\hat{\mathcal{H}}&=\int d^dr\, \hat{h}(r),
\end{align*}
where
\begin{align*}
\hat{h}(r)&\equiv \frac{1}{2} \left\{ -(\partial_i \hat{\psi})^{\dagger}(A_{ij}\partial_j \hat{\psi})
+\hat{\psi} ^{\dagger} (B_{i}i\partial_i \hat{\psi}) + (B_{i}i \partial_i \hat{\psi})^{\dagger} \hat{\psi} +\hat{\psi}^{\dagger} (C\hat{\psi})  \right\}
\end{align*}
is the Hamiltonian density operator and Hermitian [i.e $\hat{h}^{\dagger}(r)=\hat{h}(r)$].
Here $(\partial_i \hat{\psi})^{\dagger}(A_{ij}\partial_j \hat{\psi})$ is the abbreviation for
$\sum_{ij\,asa's'}(\partial_i \hat{\psi}_{as})^{\dagger} \{ (A_{ij})_{as\,a's'}\partial_j \hat{\psi}_{a's'}\}$.
A similar abbreviation is used for other terms.
Therefore, in the presence of a gravitational field $\phi (r)$,\cite{lu,enemag}
The Hamiltonian density operator and the Hamiltonian of the whole system are transformed into

\begin{align}
\hat{h}_{\phi}(r)&\equiv (1+\phi(r))\hat{h}(r) \label{en},\\
\hat{\mathcal{H}}_{\phi}&\equiv \int d^d r \hat{h}_{\phi} (r) = \int d^d r \hat{\psi} ^{\dagger} H_\phi \hat{\psi}, \nonumber
\end{align}
where
\begin{align*}
H_{\phi}(r,\partial_r)&\equiv \frac{1}{2} \left[ \sum_{ij}A_{ij} \partial _i (1+\phi)\partial_j
+\sum_{i} \{ (1+\phi)B_{i}(r)i\partial_i \right. \\
&+\left. i\partial_{i}(1+\phi)B_{i}(r) \}+(1+\phi)C(r) \right].
\end{align*}
Note the scaling relation $\left. H_{\phi} \right|_{\phi=0} = \frac{1}{2} H_{BdG}$.

Now we define the energy current operator of the system with a gravitational field as follows,
\begin{align}
\hat{j}_{E\phi \, i} (r)&\equiv \frac{1}{2} \left[ \frac{1}{2}(1+\phi (r)) \left\{ (v_i\hat{\psi})^{\dagger}(2H_{\phi} \hat{\psi}) +h.c. \right\} \right. \nonumber\\
&\left. -\varepsilon_{ijk} \partial _j \left\{ (1+\phi(r))^2\hat{\Lambda}_k \right\} \right], \label{ec}
\end{align}
where
\begin{align*}
\hat{\Lambda}_i\equiv \frac{1}{8i} \varepsilon_{ijk}(v_j \hat{\psi})^{\dagger} (v_k \hat{\psi}).
\end{align*}
The last term of eq. (\ref{ec}) is indispensable for preserving the scaling law: \cite{enemag}
\begin{align}
\hat{j}_{E\phi i}(r)=(1+\phi(r))^2\hat{j}_{E i}(r), \label{sca}
\end{align}
where
\begin{align*}
\hat{j}_{E i}(r)\equiv \left.\hat{ j}_{E\phi i}(r) \right|_{\phi=0}=
\frac{1}{2}\left\{\frac{1}{2} (v_i\hat{\psi})^{\dagger}(H_{BdG} \hat{\psi}) +h.c. 
 -\varepsilon_{ijk} \partial _j \hat{\Lambda}_k \right\}.
\end{align*}
We can check that the scaling law eq. (\ref{sca}) actually holds by a straightforward calculation 
with paying attention to $H_{\phi}=(1+\phi)\frac{1}{2}H_{BdG}-\frac{i}{4}(\partial_i \phi)v_i$.

These Hamiltonian density and energy current operators eqs. (\ref{en}) and (\ref{ec}) indeed satisfy the continuity equation:
\begin{align}
\frac{\partial \hat{h}_{\phi}(r)}{\partial t} \equiv -i \left[ \hat{h}_{\phi}(r), \hat{\mathcal{H}}_{\phi} \right] =-\sum_{i} \frac{\partial \hat{j}_{\phi i}(r)}{\partial r_{i}}. \label{con}
\end{align}
We present a brief proof of the continuity equation eq. (\ref{con}) in the following.
By noting the equations $\dot{\hat{\psi}}=-i\left[ \hat{\psi},\hat{ \mathcal{H}}_{\phi} \right]=-2iH_{\phi}\hat{\psi}$ and $v_i=2iA_{ij}\partial_j -2B_i$,
we can get
\begin{align*}
\dot{\hat{h}}_{\phi}
=&(1+\phi) \{ -\frac{1}{2}(v_i\hat{\psi})^{\dagger}(\partial_i H_{\phi}\hat{\psi})
-(B_{i}\partial_i \hat{\psi})^{\dagger} (H_{\phi}\hat{\psi})
-i(C\hat{\psi})^{\dagger}(H_{\phi}\hat{\psi}) \}\\
&+h.c. \, .
\end{align*}
Note that the factor 2 of the RHS of the commutation relation shown above follows from the relation of operators $\hat{\psi}_{as}(r)=\hat{\psi}^{\dagger}_{-as}(r)$
and the PHS of the Hamiltonian $H_{\phi \,as\,a's'}=-H_{\phi \,-as\,-a's'}^{*}$.

On the other hand, by noting the equation $\partial_i(1+\phi)v_i=4i \left[ H_{\phi} - \frac{i}{2} B_i(1+\phi)\partial_i -\frac{1}{2} (1+\phi)C\right]$,
we get the equation:
\begin{align*}
\partial_i \hat{j}_{E\phi \, i} = & \frac{1}{4} \left[  \left\{ \partial_i (1+\phi (r))v_i\hat{\psi} \right\} ^{\dagger}(2H_{\phi} \hat{\psi}) 
+((1+\phi (r))v_i\hat{\psi})^{\dagger}(\partial_i 2H_{\phi} \hat{\psi}) \right] \\
&+h.c. \\
=&(1+\phi) \{ \frac{1}{2}(v_i\hat{\psi})^{\dagger}(\partial_i H_{\phi}\hat{\psi})
+(B_{i}\partial_i \hat{\psi})^{\dagger} (H_{\phi}\hat{\psi})
+i(C\hat{\psi})^{\dagger}(H_{\phi}\hat{\psi}) \}\\
&+h.c.\, .
\end{align*}
Therefore, we obtain the continuity equation eq. (\ref{con}).

Using the energy current operator eq. (\ref{ec}) satisfying the conservation law, 
we calculate the thermal Hall conductivity in the next section.


{\it Thermal Hall conductivity} ---
In this section, we calculate the thermal Hall conductivity of superconductors using the procedure which was introduced by Qin, Niu and Shi,\cite{enemag}
with a particular attention to the symmetry of the eigenfunction eq. (\ref{sym})
and compare the result with the case of normal metals and band insulators.

The thermal Hall conductivity is given as follows:
\begin{align}
\kappa^{tr}_{xy}=\kappa^{Kubo}_{xy}+\frac{2M^z_E}{TV}. \label{thc}
\end{align}
The first term is given by the usual Kubo formula,\cite{lu} and $M^z_E$ is the gravitomagnetic energy (heat) magnetization, which characterizes the circulation of the energy (heat) flow.\cite{comment}

From now on, we use a bra-ket notation:  $\langle f|\mathcal{O}|g \rangle$ means $\sum_{asa's'}\int d^d r f^*_{as} (r)\mathcal{O}_{asa's'}(r, \partial_r) g_{a's'}(r)$.

Now we introduce new operators, which are the Fourier transforms of the field operators, and expand them in the operators
$\hat{\psi}_{kn}$ and $\hat{\psi}^{\dagger}_{kn}$:
\begin{align*}
\hat{h}_{-q}&\equiv \int d^d r e^{-i(-q)r} \hat{h}(r)=\sum_{kn\,k'n'} \hat{\psi}^{\dagger}_{kn}\hat{\psi}_{k'n'} h_{-q\,kn\,k'n'}, \\
\hat{j}_{E\,q\,i}&\equiv \int d^d r e^{-iqr} \hat{j}_{Ei}(r)= \sum_{kn\,k'n'} \hat{\psi}^{\dagger}_{kn}\hat{\psi}_{k'n'}  j_{E\,q\,x\,kn\,k'n'},
\end{align*}
where 
\begin{align}
h_{-q\,kn\,k'n'}&\equiv \frac{1}{2} \langle f_{kn} | \frac{H_{BdG}e^{iqr}+e^{iqr}H_{BdG} }{2}
| f_{k'n'} \rangle + \mathcal{O}(q^2)\nonumber\\
&\qquad \qquad \qquad \qquad \qquad \qquad (as\quad q\rightarrow0), \label{hexp} \\
j_{E\,q\,j\,kn\,k'n'}&\equiv
\frac{1}{2} \left[ \frac{ \langle H_{BdG} f_{kn} | e^{-iqr} | v_j  f_{k'n'} \rangle
+\langle v_j f_{kn} | e^{-iqr} | H_{BdG}  f_{k'n'} \rangle }{2} \right. \nonumber \\
&-\frac{1}{8i} \sum_i \left\{ \langle \partial_i v_j  f_{kn} | e^{-iqr} | v_i  f_{k'n'} \rangle
+\langle v_j  f_{kn} | e^{-iqr} | \partial_i v_i  f_{k'n'} \rangle \right. \nonumber \\
& \left. \left. -  \langle \partial_i v_i  f_{kn} | e^{-iqr} | v_j  f_{k'n'} \rangle
-\langle v_i  f_{kn} | e^{-iqr} | \partial_i v_j  f_{k'n'} \rangle
\right\} \right], \label{jexp}
\end{align}
which can be obtained by noting the two identities:
$A_{ij}\partial_i e^{iqr} \partial_j =(A_{ij}\partial_i \partial_j e^{iqr}+e^{iqr} A_{ij}\partial_i \partial_j)/2 + \mathcal{O}(q^2)$ and
$e^{-iqr}B_{i}i\partial_i+i\partial_{i}B_{i}e^{-iqr}=\{ e^{-iqr}(B_{i}i\partial_i+i\partial_{i}B_{i})+(B_{i}i\partial_i+i\partial_{i}B_{i})e^{-iqr} \}/2$.
Note that the coefficients for the expansion, $h_{-q\,kn\,k'n'}$ and $j_{E\,q\,j\,kn\,k'n'}$, preserve the symmetries
\begin{align}
h_{-q\,kn\,k'n'}&=-h_{-q\,-k'-n'\,-k-n}, \nonumber \\
j_{E\,q\,j\,kn\,k'n'}&=-j_{E\,q\,j\,-k'-n'\,-k-n}, \label{symc}
\end{align}
which follow form $H_{BdG\,as\,a's'}=-H_{BdG\,-as\,-a's'}^{*}$ and $v_{i\, asa's'}=v^{*}_{i\, -as-a's'}$,
and eq. (\ref{sym}): $f_{kn \, as}(r)=f^*_{-k-n\,-as}(r)$.

The first term of eq. (\ref{thc}) is give by
\begin{align}
\kappa^{Kubo}_{xy}=\frac{1}{VT^2}\int _0^{\infty} dt e^{-0t} \langle \hat{J}_{Ey};\hat{J}_{Ex}(t) \rangle,
\end{align}
where $\hat{J}_{Ei}\equiv \left. \hat{j}_{E\,q\,i} \right|_{q=0}$, $\langle \hat{a};\hat{b} \rangle\equiv 1/\beta \int_0^{\beta}d\lambda
\langle \hat{a}(-i\lambda ) \hat{b} \rangle$,
$\hat{a}(t)\equiv e^{i\hat{\mathcal{H}}t} \hat{a}  e^{-i\hat{\mathcal{H}}t}$, and $V$ is the volume of the system.
By the formula for the four-point correlation function eq. (\ref{4p}) and the symmetry of the coefficient eq. (\ref{symc}),
we get 
\begin{align}
\kappa^{Kubo}_{xy}&=-\frac{1}{VT^2} \sum_{\stackrel{kn\,k'n' }{(kn)\neq (k'n')}} \frac{f(E_{kn})-f(E_{k'n'})}{i(E_{kn}-E_{k'n'})^2} \nonumber \\
&\qquad \qquad \qquad \quad \times J_{Ey\,kn\,k'n'}(J_{Ex\,k'n'\,kn}-J_{Ex\,-k-n\,-k'-n'}) \nonumber \\
&=-\frac{2}{VT^2} \sum_{\stackrel{kn\,k'n' }{(kn)\neq (k'n')}} \frac{f(E_{kn})-f(E_{k'n'})}{i(E_{kn}-E_{k'n'})^2}J_{Ey\,kn\,k'n'}J_{Ex\,k'n'\,kn},
\label{kubo}
\end{align}
where $J_{Ei\,kn\,k'n'}\equiv \left. j_{E\,q\,j\,kn\,k'n'} \right|_{q=0}$.
The factor $2$ in front of eq. (\ref{kubo}) is a result of the PHS. 

Moreover, by calculating in the manner similar to ref. \citen{enemagsup}, we obtain
\begin{align}
\kappa^{Kubo}_{xy}=\frac{1}{4 T V} \sum_{kn} \mathrm{Im} \Braket{ \frac{\partial u_{kn}}{\partial k_x} | \left( H_{BdG\,k}+ E_{kn} \right)^2 
| \frac{\partial u_{kn}}{\partial k_y } } f(E_{kn}), \label{kubo2}
\end{align}
where $H_{BdG\,k}\equiv e^{-ikr}H_{BdG}e^{ikr}$.

Next we calculate the gravitational magnetization $M^z_E$.
It is the solution of the differential equation
\begin{align*}
2M^z_{E}- T\frac{\partial M^z_{E}}{\partial T} =
\frac{\beta}{2i} \left.\left\{ \frac{\partial}{\partial q_x}  \langle \hat{h}_{-q};\hat{j}_{E\,q\,y} \rangle 
-\frac{\partial}{\partial q_y}  \langle \hat{h}_{-q};\hat{j}_{E\,q\,x} \rangle \right\} \right|_{q \rightarrow 0}.
\end{align*}
with a boundary condition $\lim_{T \rightarrow 0}T\frac{\partial M^z_{E}}{\partial T} =0 $.
In order to evaluate it, we also carry out a calculation similar to ref. \citen{enemagsup}, with paying attention to the last extra term of eq. (\ref{4p}) and
the symmetries eq. (\ref{symc}).
We get
\begin{align}
M^z_{E}=-\frac{1}{4} \sum_{kn} & \left[ \frac{1}{2}\mathrm{Im} \Braket{ \frac{\partial u_{kn}}{\partial k_x} | \left( H_{BdG\,k}+ E_{kn} \right)^2 
| \frac{\partial u_{kn}}{\partial k_y } } f(E_{kn}) \right.\nonumber \\
& \left. -2E^{2}_{kn} \mathrm{Im} \Braket{ \frac{\partial u_{kn}}{\partial k_x} | \frac{\partial u_{kn}}{\partial k_y }} f(E_{kn}) \right. \nonumber \\
& \left. +4\mathrm{Im} \Braket{ \frac{\partial u_{kn}}{\partial k_x} | \frac{\partial u_{kn}}{\partial k_y }}
\int_{0}^{E_{kn}} x f(x) dx  \right] . \label{mz}
\end{align}

Note that the expressions eqs. (\ref{kubo2}) and (\ref{mz}) are one half of those in the case of
normal metals or insulators [cf: eq. (23) of ref. \citen{enemag} and eq. (S97) of ref. \citen{enemagsup}].
This is
due to the factors $1/2$ of eqs. (\ref{hexp}) and (\ref{jexp}) 
[cf: eqs. (S75) and (S76) of ref. \citen{enemagsup}] and $2$ of eq. (\ref{kubo}) [cf: eq. (S60) of ref. \citen{enemagsup}],
which are caused by the last extra term of eq. (\ref{4p}) and
the relations eq. (\ref{symc}) associated with the PHS.

In the end, we obtain an expression for the thermal Hall coefficient:
\begin{align}
\kappa^{tr}_{xy}=-\frac{1}{T V}\int dE E^2\sum_{\substack{kn\\E_{kn} \leq E}} \mathrm{Im} \Braket{ \frac{\partial u_{kn}}{\partial k_x} | \frac{\partial u_{kn}}{\partial k_y }}f'(E).
\end{align}
Moreover, by using the Sommerfeld expansion, we obtain an expression in the low-temperature limit:
\begin{align}
\kappa^{tr}_{xy}=\frac{C_1(0)}{2}\frac{\pi T}{6},
\end{align}
where $C_{1} (E)$ is the TKNN number,
which is an integer when the energy $E$ lies in the energy gap,\cite{tknn,kh}
and it is given by,
$C_{1}(E) \equiv \sum_{n} \int \frac{d^2 k}{\pi} \mathrm{Im} \Braket{ \frac{\partial u_{kn}}{\partial k_x} | \frac{\partial u_{kn}}{\partial k_y }}  \Theta(E-E_{kn}),$
where $\Theta(x)$ is the Heaviside step function.

It is notable that the quantization value $\frac{1}{2}\frac{\pi T}{6}C_1(0)$ is exactly one half of the value of the Chern insulator (or the IQHE state).
In the case of spineless chiral p-wave superconductors,
the TKNN number $C_1(0)$ is equal to $\pm 1$\cite{vol}, 
and thus, this result is in agreement with the result obtained from the Ising CFT with central charge $c = 1/2$
for the edge state.\cite{rg,nomura}

{\it Summary} ---
We have demonstrated that the thermal Hall conductivity of 2D TSCs with broken time reversal symmetry is quantized: $\kappa_{xy}=\frac{C_1}{2}\frac{\pi T}{6}$,
where $C_1$ is the TKNN number of the BdG Hamiltonian.
Our approach, solely, relies on bulk calculations, without referring to the Majorana edge theory.
This value $\frac{C_1}{2}\frac{\pi T}{6}$ is one half of the thermal Hall conductivity in the case of the Chern insulator (or the IQHE state).
In the derivation of this result, the PHS, which BdG Hamiltonians generally possess, plays an important role.
Our result is in perfect agreement with that obtained from the Ising CFT which describes the edge state of 2D TSCs.

We are thankful to Ken Shiozaki for helpful discussions.
This work is supported by the Grant-in-Aids for Scientific
Research from MEXT of Japan (Grants No. 23102714 and No. 23540406).


\begin{thebibliography}{99}
\bibitem{comment2} In ref. \citen{nomura}, it is shown that the thermal Hall conductivity of spinfull Majorana fermions on the surface of a 3D TSC is half-qunaized and associated with the TKNN number.
Moreover, this model is related to that of spinless $p+ip$ superconductors
via a unitary transformation:
$c\equiv (\psi_{\uparrow}+i \psi_{\downarrow}) / \sqrt{2}, c^{\dagger}\equiv (\psi_{\uparrow} -i \psi_{\downarrow}) / \sqrt{2}$.
\bibitem{comment} The chemical potential of the Bogoliubov quasiparticles is measured from the Fermi level, and set equal to $0$.
Thus, the energy magnetization is equivalent to the heat magnetization.
\end{thebibliography}
\end{document}